\documentclass[sigconf,natbib=true,anonymous=false]{acmart}
\usepackage{graphicx}
\usepackage{tcolorbox}
\usepackage{mathtools}
\usepackage{balance}

\def\showcomments{}
\ifdef{\showcomments}
{

}

\AtBeginDocument{%
  }

\copyrightyear{2025}
\acmYear{2025}
\setcopyright{acmlicensed}\acmConference[SIGIR '25]{Proceedings of the 48th International ACM SIGIR Conference on Research and Development in Information Retrieval}{July 13--18, 2025}{Padua, Italy}
\acmBooktitle{Proceedings of the 48th International ACM SIGIR Conference on Research and Development in Information Retrieval (SIGIR '25), July 13--18, 2025, Padua, Italy}
\acmDOI{10.1145/3726302.3730117}
\acmISBN{979-8-4007-1592-1/2025/07}

\begin{document}

%%
%% The "title" command has an optional parameter,
%% allowing the author to define a "short title" to be used in page headers.
\title{X-Cross: Dynamic Integration of Language Models for Cross-Domain Sequential Recommendation}

%%
%% The "author" command and its associated commands are used to define
%% the authors and their affiliations.
%% Of note is the shared affiliation of the first two authors, and the
%% "authornote" and "authornotemark" commands
%% used to denote shared contribution to the research.

\author{Guy Hadad}
\authornote{Corresponding author}
\email{guy_hadad@hotmail.com}
\affiliation{%
  \institution{Ben-Gurion University of the Negev}
  \city{Beer Sheva}
  \country{Israel}
}

\author{Haggai Roitman}
\authornote{Haggai Roitman was still affiliated with eBay during this work.}
\email{haggair@gmail.com}
\affiliation{%
  \institution{Ben-Gurion University of the Negev}
  \city{Beer Sheva}
  \country{Israel}
}

\author{Yotam Eshel}
\email{yeshel@ebay.com}
\affiliation{%
  \institution{eBay}
  \city{Netanya}
  \country{Israel}
}

\author{Bracha Shapira}
\email{bracha.shapira@gmail.com}
\affiliation{%
  \institution{Ben-Gurion University of the Negev}
  \city{Beer Sheva}
  \country{Israel}
}

\author{Lior Rokach}
\email{liorrk@bgu.ac.il}
\affiliation{%
  \institution{Ben-Gurion University of the Negev}
  \city{Beer Sheva}
  \country{Israel}
}

% \author{Guy Hadad, Haggai Roitman, Bracha Shapira, Lior Rokach}\authornote{Haggai Roitman was still affiliated with eBay during this work.}

% \email{{guyhada,haggair,bshapira,liorrk}@post.bgu.ac.il}
% \affiliation{%
%   \institution{Ben-Gurion University of the Negev}
%   \city{Be'er Sheva}
%   \country{Israel}
% }

% % \author{Haggai Roitman}
% % \email{haggair@post.bgu.ac.il}
% % \affiliation{%
% %   \institution{Ben-Gurion University of the Negev}
% %   \city{Be'er Sheva}
% %   \country{Israel}
% % }

% \author{Yotam Eshel}
% \email{yeshel@ebay.com}
% \affiliation{%
%   \institution{eBay}
%   \city{Netanya}
%   \country{Israel}
% }

%%
%% By default, the full list of authors will be used in the page
%% headers. Often, this list is too long, and will overlap
%% other information printed in the page headers. This command allows
%% the author to define a more concise list
%% of authors' names for this purpose.
\renewcommand{\shortauthors}{Hadad et al.}

\begin{abstract} 
As new products are emerging daily, recommendation systems are required to quickly adapt to possible new domains without needing extensive retraining. 
This work presents ``X-Cross'' -- a novel cross-domain sequential-recommendation model that recommends products in new domains by integrating several domain-specific language models; each model is fine-tuned with low-rank adapters (LoRA). 
Given a recommendation prompt, operating layer by layer, X-Cross dynamically refines the representation of each source language model by integrating knowledge from all other models. These refined representations are propagated from one layer to the next, leveraging the activations from each domain adapter to ensure domain-specific nuances are preserved while enabling adaptability across domains. Using Amazon datasets for sequential recommendation, X-Cross achieves performance comparable to a model that is fine-tuned with LoRA, while using only 25\% of the additional parameters.  In cross-domain tasks, such as adapting from \texttt{Toys} domain to \texttt{Tools}, \texttt{Electronics} or \texttt{Sports}, X-Cross demonstrates robust performance, while 
requiring about 50\%-75\% less fine-tuning data than LoRA to make fine-tuning effective. Furthermore, X-Cross achieves significant improvement in accuracy over alternative cross-domain baselines. 
Overall, X-Cross enables scalable and adaptive cross-domain recommendations, reducing computational overhead and providing an efficient solution for data-constrained environments.
\end{abstract}



%%
%% Keywords. The author(s) should pick words that accurately describe
%% the work being presented. Separate the keywords with commas.

%% Dates
% \received{13 November 2024}
% \received[revised]{---}
% \received[accepted]{---}

%%
%% This command processes the author and affiliation and title
%% information and builds the first part of the formatted document.
\maketitle

\section{Introduction}
%\hr{The sentence sounds a bit odd...} \gh{Better?} 
%\bs{the paper is about CD for sequentila recommendation- I am not sure that it is indeed only for sequential recommendation, Anyway, in the motivation, you talk about CD in general - not only for sequential. If it is only for sequential - it should be clarified with the specific challenges of sequential CD recommendation} \gh{I think the sequential framework is only a specific setting for our method, but we are not limiting for that.}

%In a sequential recommendation setting, the recommendation system is provided with the user's interactions history (e.g., clicked items) and a common goal is to predict the next item the user will interact with~\cite{BOKA2024102427}.
%In this work, we study a novel cross-domain sequential recommendation setting, where historical user-item interactions are mainly observed in several source domains, while recommended items are assumed to belong to a new or scarcely represented target domain. %\hr{this is too early I moved it to the contrinutions at the end...Our primary goal is to leverage recommendation-dedicated language models that were trained with user-item interaction histories from other source domains. Hence, we wish to learn to \textbf{transfer} knowledge that was acquired in the source domains to the new target domain; while using as \textbf{minimum} as possible observed data from the target domain. Moreover, we \textbf{do not assume knowledge sharing} of users or items between domains.}

The increasing variety of products and services available across online platforms, along with the rapid emergence of new domains, underscores the growing need for sequential recommendation models %recommendation systems 
with fast and efficient cross-domain adaptation. Traditional sequential recommendation models, often tailored to specific content areas, struggle to generalize effectively as new domains emerge at an unprecedented pace. As a motivating example, online marketplaces frequently introduce new and particular product categories, such as ``vintage collectibles'', ``sustainable fashion'', ``personalized DIY kits'' or ``specialized home automation devices''. These emerging sub-domains challenge conventional models trained on broader categories like ``consumer electronics'' or ``home goods'', since it requires nuanced understanding to capture users' preferences in such niche areas. As large language models (LLMs) continue to develop in scope and expressive power, they offer exciting potential for cross-domain applications by leveraging their vast world knowledge to quickly adapt to new domains and tasks \citep{zhao2023survey}. This capacity to operate across numerous and increasingly specific domains positions LLMs as powerful tools for recommendation tasks. %, \hr{I don't understand what purpose this sentence serve...} \gh{I want to say that a good cross-domain ability makes it possible to enlarge the number of different domains. For example, Sports can transform to more ``high resolutio'' domains like football, basketball, tennis, etc}enabling what we refer to as “higher-resolution” recommendations—recommendations that are more precise and effective because they account for a broader range of domains.

However, effective cross-domain recommendation with LLMs presents significant challenges. The large size and complexity of state-of-the-art language models make them resource-intensive to train even with parameter-efficient fine-tuning (PEFT) techniques. While PEFT, such as Low-Rank Adaptation (LoRA) \citep{hu2022lora}, has reduced the need for extensive retraining, the parameter requirements for adapting large models across multiple domains remains high, often surpassing the computational budgets of many real-world applications. A key challenge is ensuring that cross-domain recommendation tasks capture domain-specific granularity while remaining adaptable across domains. Current methods \citep{buehler2024x, xu2024meteora} rely heavily on the activations of pre-trained weights, which may lack the capacity to generate sufficiently distinctive representations for relatively similar domains. Consequently, these approaches might not adequately capture the unique nuances of each domain.

The need for adaptability in cross-domain recommendation is intensified by the data demands of LLMs. These models typically require large, domain-specific datasets to achieve optimal performance, which is not always available -- especially in newly emerging domains where data is scarce or when domain-nuanced recommendations are required. %\hr{I find next sentence to repretitive and unclear at this stage} Moreover, the nature of cross-domain tasks, which may lack enough qualitative labels for integration decisions, necessitates a dynamic approach that adapts both per sample and per layer.  
To the best of our knowledge, such an approach, which introduces finer-grained adaptability across the model, has not yet been explored in existing research. 

% \bs{this paragraph is hard to read at this point, the reader does not have enough information to understand it,  and the next paragraph is quite a repetition and is much clearer -as you start by introducing the idea of the system before going into details- I would merge these two paragraphs and make sure it is clear enough} 

The goal of our work is, therefore, to leverage language models that were trained with user-item interaction histories observed in other (source) domains for sequential recommendation in a new (target) domain. Hence, we wish to learn to \textbf{transfer knowledge} that was acquired in source domains to the new target domain; %We further wish to achieve such transferability 
this, while using as \textbf{minimum} as possible observed data from the target domain for training. Moreover, we \textbf{do not assume knowledge sharing} of users or items between domains.

Trying to address the aforementioned challenges, we introduce the ``X-Cross'' model -- a novel dynamic integration model for cross-domain adaptability that learns to transfer knowledge from several source domain language-models to a new target domain. %``X-Cross''adaptively scales network layer-by-layer for each sample input.
Given a set of two or more LoRA fine-tuned source domain language models, for each (recommendation prompt) input, X-Cross operates layer-by-layer and computes weights, integrating the strengths of multiple domain-specific models into refined representations. These refined representations are propagated from one layer to another, ensuring domain-specific nuances are preserved while enabling cross-domain adaptability. By leveraging activations from each source domain (LoRA) adapter, the dynamic integration mechanism creates representations that are both granular and adaptable, offering a practical solution. % to these challenges. %\hr{this doesn't belong here also reppetitive already...}Developing models that can perform well with limited data while maintaining efficient cross-domain adaptability is therefore critical for practical and scalable recommendation systems.

% \begin{figure}[h]
%     \centering
%     \includegraphics[width=0.4\textwidth]{example.pdf}
%     \caption{Graphical illustration of our problem.\hr{This figure is unclear, not to mention you don't even refer it in the text...so what purpose it serves right now?} \gh{this is a initial draft, i want to add a illustration for the problem. Do you have a another idea for template for the problem?}}
%     \label{fig:prompt}
% \end{figure}
% \bs{in addition to enhancing the figure - and referring to it in the text- explaining what you show in the figure,  you need to add a comprehensive title - that explains the figure with a stand alone explanation}

%\hr{repetitive...}In this work, we propose a novel approach to cross-domain sequence recommendation that leverages multiple language models, each fine-tuned on specific domains, which are subsequently integrated into a unified system optimized for cross-domain recommendations. Our method introduces a dynamic integration mechanism that constructs layer-wise embeddings for each model, generating new representations that retain domain-specific knowledge while enhancing cross-domain adaptability. 
Such a layer-wise dynamic integration eliminates the need to retrain or modify the original source domain LoRA adapters, %\hr{it is only clear to you right now that you are using or comparing LoRA adapters in this work...the reader at this stage has no such prior knowledge. You need to start with some motivation about why LoRA is needed at all.} \gh{More than the paragraph above about PEFT and LoRA?}, 
allowing our model to achieve performance comparable to newly trained LoRA adapters using just 25\% of the parameters required by LoRA.  %\hr{additional to what?} \gh{Better?}
%Moreover, a notable advantage of X-Cross model is its efficiency with limited training data. 
Moreover, X-Cross achieves a competitive and sometime even better recommendation quality to a model that is fine-tuned with LoRA while using %only Our experiments, which involve adapting between domains such as toys, tools, electronics, and sports, show that X-Cross model can surpass zero-shot performance with 
50-75\% less training data.  %than LoRA requires for fine-tuning. \bs{not clear}
We further demonstrate that X-Cross performs better than other alternative language-model based cross-domain recommenders, including alternatives that utilize state-of-the-art mixture-of-LoRA methods. %Furthermore, in comparison with alternative techniques using mixtures of LoRA adapters, X-Cross model yields substantially improved results, underscoring the robustness and scalability of its dynamic integration capabilities. \bs{don't you want to say that you are better than other cross domain solutions}

Overall, our work contributes to developing a parameter-efficient, data-friendly, and adaptable solution for cross-domain sequential recommendation, offering a scalable pathway for recommendation systems in data-limited and rapidly evolving domains.

\section{Related Work}
We categorize previous related works into three areas of research: language models for sequential recommendation, cross-domain recommendation, and methods employing mixture-of-LoRAs for efficient model fine-tuning. Below, we briefly review these works and highlight their limitations in comparison to our approach.

\subsection{Language Models for Sequential-Recommendation}
The sequential recommendation task has been extensively studied~\cite{BOKA2024102427}, with many previous works commonly model user-item interaction sequences using Transformer-models; having sequence items represented either by their identities (IDs) (e.g., SASRec~\cite{kang2018self}, BERT4Rec~\cite{sun2019bert4rec}, SSE-PT~\cite{wu2020sse}) or by their attributes (e.g., FDSA~\citep{zhang2019feature}, Trans2D~\cite{trans2d}). The rise of large language models (LLMs) in recent years has revolutionized natural language processing (NLP), and specifically their applications to recommendation systems~\cite{liu2023pre, wu2024survey,llm4recsurvey2024} (LLM4Rec). For sequential recommendation tasks, language models have been utilized so far in two primary ways. Firstly, language models have been utilized for feature encoding and augmentation, providing rich context for downstream sequential recommendation pipelines~\cite{llm4recsurvey2024}. Secondly, previous works have utilized the in-context learning and instruction following capabilities of LLMs to provide recommendations directly to users. To this end, given user history described in a textual form (prompt), language models were instructed either to select the next item from a list of candidates (e.g., P5~\cite{geng2023P5}, InstructRec~\cite{zhang2023instructrec}, RecRanker~\cite{luo2024recranker}) or generate a template that represents such an item (e.g., LlamaRec~\cite{yue2023llamarec}, RecPrompt~\cite{Liu_2024RecPrompt}) which is then used to score and rank actual items. In this work, we also utilize a language model for sequential-recommendation, where we cast the task as a multi-choice problem, converting its outputs into scores for ranking candidates for user's next item.   

\subsection{Cross-Domain Recommendation}
The cross-domain recommendation task aims to transfer knowledge between domains to improve performance, particularly in scenarios where the target domain suffers from limited or sparse training data~\citep{zang2022survey}. Language models hold significant potential for such tasks due to their vast world knowledge and advanced reasoning capabilities, which are essential for the generalization required in cross-domain recommendation settings~\citep{tang2023one,chen2024large}.

% To the best of our knowledge, the cross-domain recommendation task has been primarily studied in scenarios where source and target domains \textbf{share overlapping users or items} \citep{yuan2020parameter, wu2020learning, bi2020heterogeneous, chen2021user, shen2024exploring, tang2023one, li2023text}. Among such previous works, %\gh{These works are not assuming that}, 
Among the most relevant works to ours, ZESRec~\citep{ding2021zero} has applied BERT \citep{devlin2019bert} to create semantic representations for zero-shot cross-domain recommendations. %\gh{ZESRec require the item the full descriptions and not only the titles} \gh{according to one paper: ZESRec requires that the source and target domains are closely related.} 
TransRec~\cite{fu2024exploring} has explored adapter-tuning for transferable recommendations. %\gh{it is worth mentioning but not relevant to our task specifically} 
UniSRec~\citep{hou2022towards} has addressed cross-domain sequential recommendation by utilizing a Mixture-of-Experts module to integrate the BERT representations into the recommendation task. VQ-Rec~\citep{hou2023learning} has adapted textual embeddings generated by pre-trained language models by leveraging vector-quantization. RecFormer~\cite{li2023text} has utilized language representations to model user preferences and item features, enabling effective next-item prediction, particularly in low-resource and cold-start scenarios.

%We focus on cross-domain settings that \textbf{do not rely on any overlap between users or items}, aiming to develop a more generalized and less data-dependent framework for cross-domain recommendation tasks.
% Recformer \citep{li2023text} is bi-directional Transformer similar to the model Longformer but with different embedding layers \gh{we already talked about the problem that the input us pretty different than ours (all the descriptions) }\hr{we need to have it as baseline as well. 1. Their motivation is similar (domain transferability) 2. they provide the code/checkpoint so you can try it \url{https://github.com/AaronHeee/RecFormer}}
%Furthermore, 
Compared to our work, existing models rely on fixed representations derived from the language model \citep{hou2022towards, hou2023learning}, which limits their ability to fully leverage the potential of the model during fine-tuning for the source domain. This is due to the inherent constraints of pre-trained language models when applied to recommendation tasks~\citep{kang2023llms}. In addition, utilizing only the final hidden state disregards the rich knowledge embedded within intermediate layers~\citep{tenney2019bert}, which may contain valuable information relevant to cross-domain recommendation tasks.

%\gh{I don't include cross domain works in general... works like "RecGURU", "C2DSR", "pi-Net" are relevant? it is pretty far settings than ours. there is also works like "TransRec" that talk about multi-modal feedback as well and this not so relevant to our type of work}

\subsection{Mixtures of LoRAs}
%\hr{SMEAR https://arxiv.org/pdf/2306.03745}
With the rise of parameter-efficient fine-tuning (PEFT) methods \citep{fu2023effectiveness, ding2023parameter}, driven by the increasing scale of models and datasets, LoRA \citep{hu2022lora} and its variants \citep{dettmers2024qlora, liu2024dora, kalajdzievski2023rank} have emerged as a popular approach for fine-tuning not only large language models but also other large-scale neural networks across diverse applications~\citep{zhang2023adding, liu2024visual}. By significantly reducing the number of trainable parameters, LoRA enables efficient adaptation without modifying the pre-trained model weights, making fine-tuning more computationally feasible. The integration of several models has gained a significant traction in various generative AI applications~\citep{yadav2023resolving, yadav2024ties}; with one of the most popular approaches being the integration of several LoRA adapters, each fine-tuned on a different task~\citep{feng2024mixture, prabhakar2024lora}. Previous works  have explored various strategies for integrating adapters in machine-learning models. Some works focus on integrating adapters exclusively during the inference phase to enhance model performance without additional training overhead \citep{prabhakar2024lora}. Others extend this approach by integrating adapters during training, often incorporating Mixture-of-Experts \citep{fedus2022review, fedus2022switch, zhou2022mixture} (MoE) techniques to better optimize the models for specific tasks \citep{buehler2024x, xu2024meteora}.
Furthermore, integration mechanisms have been proposed, with most approaches relying on pre-trained model representations as the foundation for the integration process. While some methods require domain labels during the training process \citep{xu2024meteora, feng2024mixture}, they still often face challenges in cross-domain tasks where true labels may be ambiguous or unavailable~\citep{xu2024meteora}. These approaches usually assume distinct and well-separated domains during training, which is rarely the case in real-world recommendation tasks where domain boundaries can be vague. %\hr{citation} \gh{This is a simple statement, when i want to recommend on toys i don't have labels which domains to choose}. 
Moreover, these methods can struggle when the pre-trained model representations are not sufficiently distinctive, as is often the case in recommendation tasks~\citep{kang2023llms}.

\section{Recommendation Framework}
In this section we describe our proposed solution for cross-domain sequential recommendation. We begin by providing the basic background required to understand the cross-domain sequential recommendation task and how we leverage a language-model to solve it. We then present the X-Cross model which integrates several source domain language-models to recommend items in a new target domain. We describe its four different stages and how to utilize it for cross-domain sequential recommendation. 

\subsection{Background}
%\hr{I organized everything related to background to our method in this sub-section. I further simplified notations.} 
We provide the necessary background for our model. First, we define the cross-domain sequential recommendation task. Next, we explain how we frame it as a multiple-choice problem using a language model. We then outline the model training process and introduce LoRA for domain-specific fine-tuning.

\subsubsection{Cross-Domain Sequential Recommendation Task}
Let \( U \) represent the set of users, \( I \) the set of items, and $S_u = \left(i_1, i_2, \dots, i_N \right)$ represent the (sequence) history of $N$ last items from \( I \) that user \( u \in U \) has interacted with.
Given user history \( S_u \), the sequential recommendation task is to predict the next user interaction \( i_{N+1}\in{I} \). 

In the cross-domain recommendation setting that we study in this work, items in \( I \) are assumed to belong to a new or scarcely represented target domain $D_{target}$; where we aim to leverage recommendation models that were trained using user-item interaction histories that were observed in other source domains $\mathcal{D}_{source}=\{D_1,D_2,\ldots,D_n\}$. Hence, we wish to learn to \textbf{transfer knowledge} that was acquired in the source domains to the new target domain. Moreover, we wish to achieve such transferability assuming \textbf{minimum} %as possible 
observed data from the target domain.

\subsubsection{Sequential Recommendation as a Multiple-Choice Problem}\label{sec:multichoice task}
In this work, we use a language model to recommend the next item to a user \( u \in U \), given user's interaction history \( S_u \). To this end, we first cast the recommendation task as a multiple-choice problem. %We further assume that the next item with which the user may interact should be selected from a closed set of candidate items \( C_u \). 
For any given candidate next item $i\in{I}$, the input to the language model is simply expressed in a textual form (prompt) as follows:
\[
\text{Prompt} = \left[\text{History: } S_u, \text{Candidate: } i\right]
\]

\begin{figure}[t!]
\small
\begin{tcolorbox}[colframe=white!90!black, colback=blue!3!white, coltitle=black, title=Sequential Recommendation Prompt]
\textbf{History:} \\[0.1em]
1: SE 8787MS-SP 9-in-1 Emergency Shovel Tool Kit. \\[0.1em]
2: SE CCH7-1G 7-IN-1 Green Survival Whistle. \\[0.1em]
3: Attmu 2 Pack Outdoor Survival Paracord Bracelet. \\[0.1em]
4: Garmin DeLorme Atlas \& Gazetteer Paper Maps - Arizona. \\[0.1em]
5: DANTENG Phone Cable Black\\[0.1em]
\textbf{Candidate:} \\[0.1em]
Mylar Men's Emergency Thermal Blankets (10 Pack)
\end{tcolorbox}
\caption{Example prompt for sequential recommendation task having a user history with 5 items and a candidate item.}
\label{fig:prompt}
\end{figure}

% \begin{figure}[t!]
% \centering
% \small
% \begin{tcolorbox}[colframe=black,colback=white,title=Sequential Recommendation Prompt]
% \textbf{History:} \\[0.3em]
% 1: SE 8787MS-SP 9-in-1 Emergency Shovel Tool Kit. \\[0.3em]
% 2: SE CCH7-1G 7-IN-1 Green Survival Whistle. \\[0.3em]
% 3: Attmu 2 Pack Outdoor Survival Paracord Bracelet. \\[0.3em]
% 4: Garmin DeLorme Atlas \& Gazetteer Paper Maps - Arizona. \\[0.3em]
% 5: DANTENG Phone Cable Black\\[0.3em]
% \textbf{Candidate:} \\[0.3em]
% Mylar Men's Emergency Thermal Blankets (10 Pack)
% \end{tcolorbox}
% \caption{Example prompt for sequential recommendation task having a user history with 5 items and a candidate item.}
% \label{fig:prompt}
% \end{figure}

Figure~\ref{fig:prompt} illustrates an example prompt with a history that contains five items and a candidate item. 

%Assuming a generative form for solving the problem, 
We now define $score(S_u,i)$ as the model ``confidence'' that item $i$'s text is likely to follow the textual representation of the user history $S_u$ in the prompt. Using the language-model as an \textbf{encoder}, in this work, we obtain such a score by simply pooling the representation of the last hidden layer of the language model  and apply a simple scoring ``head'' over it (see details in Section~\ref{sec:scoring}). Given a set of candidate items in $I$, using such a prompting (and scoring) approach, allows us to score any candidate item  and select the next item as the one with the highest score. %\ye{this paragraph needs rephrasing. We write 'Assuming a generative form for solving the problem' but we do not propose a generative form to solve the problem. We as well write 'Using the language-model as an encoder'. I'm confused. Are we using encoder LMs (e.g. Bert/Deberta) or decoder LMs (e.g. GPT)?}

\subsubsection{Model Training}\label{sec:modal training}
Following a common methodology in training recommender-systems~\citep{ren2024representation, yuan2020parameter, yuan2021one}, %\gh{I add the papers that do the splitting like me. These the papers I need to cite here?},
we train the (language) model by sampling several negative items $I_{neg}$ for each true item $i$ with which a given user $u$ has interacted. %\hr{Comment: state in the experiments how negative samples are obtained} \gh{I write about that in the experimental setup, add it also here? it is more about implementation and less about methodology}
%The candidate pool \( C_u \) includes the true next item \( s_N \) (positive) and \( m-1 \) negative samples \( \{c_1, \dots, c_{m-1}\} \), sampled based on item popularity. 
For each true candidate item $i\in{I}$, the model is, therefore,  trained to maximize $score(S_u,i)$ using the negative log-likelihood loss:
\begin{equation}\label{loss function}
\mathcal{L}_{i} = -\log \left( \frac{\exp(score(S_u,i))}{\sum_{i'\in\{i\}\cup {I_{neg}}} \exp(score(S_u,i'))} \right).
\end{equation} %\bs{what about the model's inference?}\hr{that's trivial, we described how we already score documents...}

%where $y_i$ equals to $1$ only for item $i$ and $0$ for the negative items. 

\subsubsection{LoRA for Domain Specific Fine-Tuning}
Low-Rank Adaptation (LoRA) \citep{hu2022lora} is a method designed to efficiently fine-tune large pre-trained models.  To this end, LoRA introduces additional trainable low-rank matrices into specific layers, based on the hypothesis that the updates required for fine-tuning reside in a low-dimensional subspace. Therefore, instead of updating all the parameters of the pre-trained model, LoRA freezes the original model weights %( \mathbf{W} \in \mathbb{R}^{d \times k} \) 
and introduces a learnable adjustment:
\begin{equation}\label{LoRA}
\mathbf{W}_{\text{LoRA}} = \mathbf{W} + \alpha \cdot \mathbf{A} \mathbf{B},
\end{equation} 
where:
\begin{itemize}
    \item \( \mathbf{W} \in \mathbb{R}^{d \times k} \) represents the frozen pre-trained weights, with \( d \) denoting the input dimension and \( k \) the output dimension.
    \item \( \mathbf{A} \in \mathbb{R}^{d \times r} \) and \( \mathbf{B} \in \mathbb{R}^{r \times k} \) are trainable low-rank matrices, where \( r \ll \min(d, k) \), representing the adaptation rank.
    \item \( \alpha \in \mathbb{R} \) %is a scalar scaling factor 
    controls the adaptation magnitude.
\end{itemize}

In this work, for each source domain \( D\in\mathcal{D}_{source} \), we train a unique LoRA adapter to capture domain-specific patterns.

\subsection{The X-Cross Model}
%\hr{I organized the section for better flow, also simplified notations and made them more consistent.}
The X-Cross model (hereinafter termed ``X-Cross'' for short) learns to recommend items in a new target domain by integrating multiple pre-trained language models, with each model that was previously adapted with LoRA for a specific source domain. X-Cross dynamically combines activations from the LoRA-enhanced ``encoders'' across all source domains, allowing the model to leverage domain-specific knowledge while adapting flexibly to the input context. 

\begin{figure}[t!]
    \centering
    \includegraphics[width=0.5\textwidth]{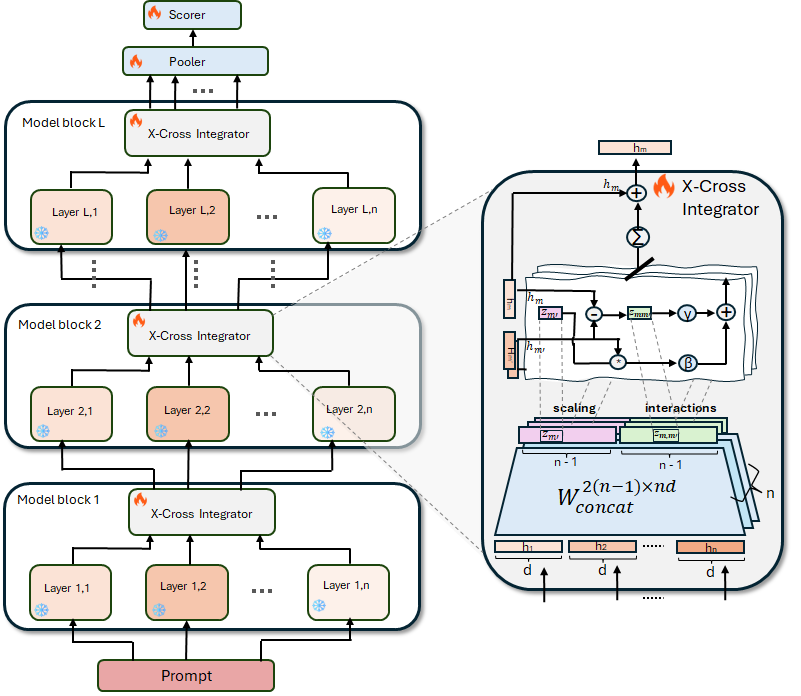}
    \caption{X-Cross model architecture. Each source domain language model is implemented with several Transformer (vertical) layers. On the left side: at each layer, the ``hot''-trainable integrator receives activations from the ``frozen'' layers and then passes the integrated representations to the next layer. On the right side: a ``zoom-in'' into an X-Cross integrator located at one of the network layers.} %which learns the ``integration policy'' between the various source domains. } 
    %\bs{can you mark or say explicitly how are the source domains represented  }
    \label{fig:methodology}
\end{figure} 

Figure~\ref{fig:methodology} now illustrates the architecture of X-Cross. As a preliminary step, we assume the availability of $n\geq{2}$ source domain (language) models, assuming each model is fine-tuned with its own dedicated LoRA adapter. The source models are visually represented in a vertical arrangement within the figure. X-Cross pools information from all source domain models by introducing a dedicated integrator at each (network) layer, enabling the computation of scaling factors (weights) dynamically for each source domain. This per-layer integration design captures the varying importance of layers in recommendation tasks, where different layers contribute uniquely to relevance signals. By dynamically adjusting the scaling factors at each layer, X-Cross ensures that the integrated representation effectively balances shared and domain-specific knowledge. Our experimental results (see Section~\ref{sec:ablation}) further validate the necessity of per-layer integration, demonstrating its critical role in achieving effective cross-domain recommendations. Overall, X-Cross implementation includes four main stages. The first three  stages are applied at each layer-level for each source domain, while the final stage is applied on the outputs of the last layers of all source domains. For simplicity of presentation, the detailed stages describe representation calculations that are performed over every sequence input token (i.e., we exclude the sequence-dimension). %\hr{Guy - check} \gh{do you think it is obvious enough to the reader that it is related to the dimensions?} \gh{Ok, I see you write it later explicitly}

\subsubsection{Stage 1: Concatenation of source domain representations} 
For each source domain $D_m$ ($1\leq{m}\leq{n}$), let $\mathbf{h}^{(l)}_{m}$ denote the output of the LoRA-enhanced encoder for that domain at layer $l$ ($1\leq{l}\leq{L}$) of the network, which is obtained as follows:
\vspace{-0.01in}
\begin{equation}\label{per domain LoRA}
\mathbf{h}^{(l)}_{m} = \left[\mathbf{W}^{(l)}_{m} + \mathbf{A}^{(l)}_{m}\mathbf{B}^{(l)}_{m}\right]\cdot \mathbf{x}^{(l)}_{m}, 
\end{equation}
where \(\mathbf{W}^{(l)}_{m} \in \mathbb{R}^{d \times k}\) denotes the \textbf{frozen} pre-trained weights matrix of the $m$-th encoder at layer $l$; \(\mathbf{A}^{(l)}_{m} \in \mathbb{R}^{d \times r}\) and \(\mathbf{B}^{(l)}_{m} \in \mathbb{R}^{r \times k}\) are the LoRA weights specific to the encoder; and \(\mathbf{x}^{(l)}_{m} \in \mathbb{R}^{k}\) represents the input to layer $l$. Here we note that, %at this stage, 
we assume that the fine-tuned LoRA weights (i.e., $\mathbf{A}^{(l)}_{m}$ and $\mathbf{B}^{(l)}_{m}$) are \textbf{frozen} as well.  

In contrast to previous models \citep{xu2024meteora, buehler2024x} that have applied integration directly on activations from pre-trained weights (i.e., \(\mathbf{Wx}\)), our approach routes the adapted activations (i.e., \((\mathbf{W} + \mathbf{AB})\mathbf{x}\)) produced after domain-specific LoRA adapters. This adjustment addresses a known limitation in recommendation tasks, where pre-trained model activations often lack sufficient knowledge about recommendation-specific data~\citep{kang2023llms}. By leveraging the enriched activations from LoRA adapters, X-Cross dynamically adjusts weights across representations from different domains without requiring explicit supervision for the integrator. Instead, the weights are computed adaptively for each input, guided solely by the supervision provided by the label of the sequential recommendation task in the target domain. This approach enables label-free integration of domain representations, enhancing flexibility and performance in cross-domain recommendations. %\ye{this two sentence are hard for me to understand}. \hr{I agree with Yotam here, it reads to me like we don't need labels at all, which is incorrect, we do use some amount of target domain examples...Guy please rephrase.} \gh{better now?}

At this stage we obtain the concatenated representation across all $n$ encoders:
\[
\mathbf{h}_{\text{concat}}^{(l)} = \big\lBrack \mathbf{h}_{1}^{(l)}; \mathbf{h}_{2}^{(l)}; \dots; \mathbf{h}_{n}^{(l)}\big\rBrack,
\]
where $\big\lBrack\cdot; \cdot\big\rBrack$ represents the concatenation operation along the feature dimension. Each $\mathbf{h}_{m}^{(l)} \in \mathbb{R}^{d}$ contributes to the final representation, resulting in $\mathbf{h}_{\text{concat}}^{(l)} \in \mathbb{R}^{n \cdot d}$.

\subsubsection{Stage 2: Dynamic scaling} 
%\ye{I am a bit confused about this entire subsection... I didn't get what are we trying to say here}

%\ye{Only after reading the next subsection I get what this is. I think we should find a new term for 'Weights computation'. In previous sections we talked about 'scaling factors' and 'dynamic scaling'. We should keep the terminology here... Weights is somehow confusing. }

At this stage, the concatenated representation \( \mathbf{h}_{\text{concat}}^{(l)} \) undergoes a trainable linear transformation to compute the weights for domain-specific layer representations. %For each domain, we now calculate weights to account for both \textbf{direct domain-specific} contributions (overall \( n-1 \) such weights) and \textbf{inter-domain interactions} with every other domain-specific layer (overall \( n-1 \) additional weights), resulting in \( 2(n-1) \) weights per source domain. Given \( n \) total source domains, this computation scales to \( 2n(n-1) \) weights in total. \hr{there is a mistake in next sentence...}These weights are subsequently used to construct the new layer representation $\mathbf{z}^{(l)}$. 
Formally, the weights are computed as:
\begin{equation}\label{new layer representation}
\mathbf{z}^{(l)} = \mathbf{W}^{(l)}_{\text{concat}}\cdot \mathbf{h}_{\text{concat}}^{(l)},
\end{equation}
where \( \mathbf{W}^{(l)}_{\text{concat}} \in \mathbb{R}^{2n(n-1) \times (n \cdot d)} \) is a trainable weight matrix. Therefore, the output \( \mathbf{z}^{(l)} \in \mathbb{R}^{2n(n-1)} \) contains for each domain: 1) n-1 \textbf{domain-specific} scaling factors which modulate the contribution of other domains to the integrated representation for that domain; 2) n-1 \textbf{interaction terms} per each \textbf{pair} of that domain with another domain. Hence, in total, $\mathbf{z}^{(l)}$ contains \(2n(n-1)\) learnable weights. These weights in $\mathbf{z}^{(l)}$ enable the model in the next stage to dynamically balance the contributions from each source domain while capturing complex inter-domain dependencies, resulting in richer and more nuanced integrated representations.

Unlike conventional methods that rely on softmax to produce only positive scaling factors~\citep{xu2024meteora, zhou2022mixture}, our approach supports both positive and negative scaling. This flexibility allows X-Cross to \textbf{suppress irrelevant domain contributions}. This enhances X-Cross's ability to focus on meaningful features for the recommendation task. By allowing negative scaling, X-Cross can effectively downscale or even “move away” from less relevant domains when ``constructing'' the integrated representation. This capability is particularly advantageous in scenarios where certain domain adapters introduce noise or irrelevant information, ensuring the final representation prioritizes the most relevant features. %\ye{I think this paragraph needs going over this again. I think this is a bit hand wavy. having negative weights sounds like it would account for negative correlations rather than suppress irrelevant features}\hr{I made the text bold to emphasize the meaning}

%\hr{I commented out this part...we are talking about this in the next subsection anyway..}Additionally, X-Cross introduces inter-domain interactions as a critical dimension in the scaling process. Rather than solely considering the individual contributions of domain adapters (e.g., \texttt{Tools} or \texttt{Electronics}), the integrated representation also accounts for interactions between them by considering the representation differences $\left(\mathbf{h}^{(l)}_{m} - \mathbf{h}^{(l)}_{m'}\right)$ for any source domain pair $D_m$ and $D_{m'}$\ye{could this h\_m-h\_m' belong in the next bullet (3.2.3)?}.  This approach enriches the embedding space by capturing nuanced relationships between domains, which are crucial for addressing complex recommendation tasks effectively.

\subsubsection{Stage 3: Representation Refinement and Integration:}
%\ye{I think 3.2.2 should come after this subsection since here we understand what these z's are. Then 3.2.2 will  explain how are they calculated. wdyt?}\hr{No, it is OK as is. z is need before we apply (5)}
This stage ensures that the integrated representation of each domain $D_m$ (i.e., $\mathbf{h}^{(l)}_{m}$) would ``blend'' knowledge from all domains while dynamically adapting to the context of each sample. The weights $\mathbf{z}^{(l)}$ modulate both the direct contributions of domain-specific outputs and their interactions with other domains. We next detail how these weights are applied to construct the final  representation of each domain.

For each domain-specific representation $\mathbf{h}^{(l)}_{m}$, we compute an integrated refined representation, denoted as \(\tilde{\mathbf{h}}^{(l)}_{m}\). This representation incorporates contributions from other domains while excluding the current domain $D_m$, and is calculated as follows:
\vspace{-0.01in}
\begin{equation}\label{domain intergation}
\tilde{\mathbf{h}}^{(l)}_{m} = \mathbf{h}^{(l)}_{m} + \sum_{m' \neq m} \left( \beta \cdot \mathbf{z}^{(l,m)}_{[m']} \cdot\mathbf{h}^{(l)}_{m'} + \gamma \cdot\mathbf{z}^{(l,m)}_{[m,m']}\cdot \left(\mathbf{h}^{(l)}_{m} - \mathbf{h}^{(l)}_{m'}\right) \right)
\end{equation}
 %\hr{Guy - check next paragraph and Eq 5} \gh{Looking good, maybe it will demand the reader to make a little effort to understand equation...}
Here, the term \(\beta \cdot \mathbf{z}^{(l,m)}_{[m']} \cdot\mathbf{h}^{(l)}_{m'}\) integrates the \textbf{direct influence} of other domains \(D_{m'}\) into the refined representation of domain \(D_m\), where $\mathbf{z}^{(l,m)}_{[m']}$ is the corresponding scaling factor for domain \(D_{m'}\). The term \(\gamma \cdot\mathbf{z}^{(l,m)}_{[m,m']}\cdot \left(\mathbf{h}^{(l)}_{m} - \mathbf{h}^{(l)}_{m'}\right)\) further captures the \textbf{inter-domain interactions} by measuring the differences between domain representations, enriching the unified representation for the cross-domain task. The term $\mathbf{z}^{(l,m)}_{[m,m']}$ further denotes the relative weight of the inter-domain interaction between the two domains. 

We note that, the summation in Eq~\ref{domain intergation} excludes domain $D_m$ itself, ensuring that the refined representation focuses on its interactions with the other \(n-1\) domains. The scalars \(\beta\) and \(\gamma\) are hyperparameters that control the contribution strength of the two terms.

The residual connection allows each domain-specific representation to retain its original characteristics by avoiding incorporation of superfluous
 knowledge, reducing noise and preserving key domain knowledge. %``integrity''.\hr{I don't like the term ``integrity''...please explain what you meant here?}

The adaptive integration mechanism operates iteratively at each layer $1\leq{l}\leq{L}$, enabling the model to dynamically balance shared and domain-specific knowledge. Through this progressive refinement, the model captures complex inter-domain relationships, producing robust integrated representations. 

\subsubsection{Stage 4: Final weighted summation:} After processing through all layers, the refined domain-specific representations are aggregated into a single integrated representation. This is achieved through a weighted summation of the final layer outputs from all source domain encoders:
\begin{equation}\label{final representation}
    \mathbf{h}^{\text{final}} = \sum_{m=1}^{n} w_{m}\cdot \tilde{\mathbf{h}}^{(L)}_{m},
\end{equation}
 %\hr{Guy - check next paragraph and Eq 6} \gh{Looking good}
where \(\tilde{\mathbf{h}}^{(L)}_{m}\) is the output of the $m$-th encoder at the last layer \(L\), and $w_{m}\in\mathbb{R}$ is the learnable weight of domain $m$.

\subsubsection{Candidate scoring}\label{sec:scoring} 
Once X-Cross has encoded the input prompt, we utilize the integrated representation \( \mathbf{h}^{\text{final}} \) ($\in\mathbb{R}^d$) to estimate the ``likelihood'' that the prompt's candidate item is the next item given user's interactions history (see again Section~\ref{sec:multichoice task}). To this end, \( \mathbf{h}^{\text{final}} \) undergoes a pooling operation to extract a compact vector that ``summarizes'' the sequence information. Specifically, we employ a \textit{contextual token pooling} mechanism that focuses on the representation of the first (\texttt{[CLS]}) token in the sequence. %\hr{Guy - check}\gh{looking good} 
This pooling method is consistent with the one proposed in the DeBERTa model~\cite{he2021deberta}, which is the language model we utilize in our experiments (see more details in Section~\ref{sec:Experimental Setup}). 

Formally, the pooled representation \( \mathbf{h}_{\text{pooled}} \) is defined as follows:
\vspace{-0.03in}
\begin{equation}\label{pooling}
\mathbf{h}_{\text{pooled}} = \text{GELU}\big(\mathbf{W}_p \cdot \mathbf{h}^{\text{final}}_{\texttt{[CLS]}} + \mathbf{b}_p\big),
\end{equation}
where:
\begin{itemize}
    \item \( \mathbf{h}^{\text{final}}_{\texttt{[CLS]}}\in\mathbb{R}^d \) represents the hidden state corresponding to the first token (\texttt{[CLS]} context token). %\hr{Guy - check} \gh{looking good, better the the zero...}
    %\item \(\text{Dropout}(\cdot)\) applies a stable dropout to prevent overfitting.
    \item \(\mathbf{W}_p \in \mathbb{R}^{d \times d}\) and \(\mathbf{b}_p \in \mathbb{R}^d\) are learnable parameters. %of the pooling layer.
    \item \(\text{GELU}(\cdot)\) is the Gaussian Error Linear Unit activation function~\citep{hendrycks2016gaussian}, as used in the DeBERTa model~\cite{he2021deberta}.
\end{itemize}

Finally, we score the prompt's candidate item $i$ as the next item in the user's sequence by applying a simple scoring (regression) ``head'' over the pooled representation $\mathbf{h}_{\text{pooled}}$. For that, we implement the ``scorer'' using a simple linear layer, calculated as follows:
\vspace{-0.01in}
\begin{equation}\label{scorer}
score(S_u,i) = \mathbf{V}^{T}_c \cdot \mathbf{h}_{\text{pooled}} + b_c,
\end{equation}
where \(\mathbf{V}_c \in \mathbb{R}^{d}\) and \(b_c \in \mathbb{R}\) are the learnable weights vector and bias term of the linear layer, respectively. 

To remind, $score(S_u,i)$ represents the model's predicted score for the given prompt's candidate item $i$ and user history $S_u$ (we kindly refer the reader again to Section~\ref{sec:modal training} for model training details). Further to remind, at training time, we train the model to score the true candidate $i$ along with a sample of negative items $I_{neg}$. At inference time, given a recall-set of candidate items, we simply choose the next user item as the one with the highest score. 

%\subsubsection{Discussion: X-Cross model efficiency}
% We conclude this section with a short discussion on the efficiency of fine-tuning for cross-domain recommendation with X-Cross compared to ``traditional'' fine-tuning with LoRA. Here we note that, the only new trainable weights in X-Cross are $\mathbf{W}^{(l)}_{\text{concat}}$ per network layer $l$. Hence, the number of additional required parameters are $\mathcal{O}(n^3Ld)$. This in comparison to a requirement for full fine-tuning with LoRA for the new target domain which would require $\mathcal{O}(r(d+k)L)$ new parameters. Assuming that $d\gg{k}$ and $r\gg{n^3}$ (e.g., whenever the number of source domains being utilized is small enough), fine-tuning with X-Cross would be much more efficient. Moreover, as we shall shortly demonstrate, X-Cross further requires significantly less samples for fine-tuning on a new target domain. \gh{Do you think O notation analysis is relevant when $n$ and $r$ are always pretty small? We assume $r \leq 64$ and $n \leq 2$ or $n \approx 3$. For each layer, we get a complexity of $n \cdot d \cdot 2 \cdot n \cdot (n-1)$, while the LoRA complexity is $2 \cdot d \cdot r$. We get $n^3 - n^2 \leq r$. While $n$ is a small number, the term $n^2$ is significant. Thus, our model appears better because saying $r \gg n^3$ is quite weak, no?} \hr{I completely removed this discussion. It can make more harm then help...in the evaluation we give exact numbers.}

\section{Evaluation}
%\hr{TODO: make sure we either use past tense or present tense consistently throughout this section...}
\label{sec:Experiments}
In this section we evaluate X-Cross. We first outline the experimental setup (Section \ref{sec:Experimental Setup}). We then evaluate X-Cross over different domains and baselines, including traditional baselines, cross-domain baselines, and alternative integration strategies (Section \ref{sec:Results}). Our evaluation further emphasizes the competitive performance of X-Cross to training a new LoRA adapter. Next, we assess X-Cross efficiency by examining its performance under limited training data conditions (Section \ref{sec: Efficiency}) and the effect of the number of % We also investigate the effect of including different numbers of
layers in gradient calculations (Section \ref{sec: Layers}). We then conduct an ablation study %to determine the contribution of each component to overall performance 
(Section \ref{sec:ablation}) and conclude with an analysis of the factors driving the diverse convergence behaviors observed across datasets (Section \ref{sec: Convergence}).

\subsection{Experimental Setup}
\label{sec:Experimental Setup}
\subsubsection{Datasets.}
    We curate four datasets from the Amazon reviews corpus~\citep{mcauley2015image}, focusing on the domains of \texttt{Electronics}, \texttt{Sports}, \texttt{Tools} and \texttt{Toys}, which are commonly used for sequential recommendation tasks \citep{kang2018self, sun2019bert4rec, li2020time}. Each dataset complies with the widely adopted ``core 5'' criteria \citep{ren2024representation, wang2019neural, he2017translation}, ensuring that every user and item has at least five interactions.  Further following~\citep{bao2023tallrec}, to accommodate the input constraints of the language model in our evaluation, we limit each user's interaction history to a minimum of 5 and a maximum of 15 unique items. We further represent each item within the input prompt by its title (see again Figure~\ref{fig:prompt}).
    
    We construct user histories by randomly selecting one interaction per day from each user's activity. This approach addresses a significant limitation of the Amazon datasets (2018), which only provide timestamps at the \textbf{day level}, leading to potential inaccuracies in time-sensitive tasks like sequential recommendation \citep{hou2024bridging}. By ensuring a consistent chronological sequence, this approach helps to preserve temporal order across days, reducing ambiguity and improving the reliability of sequence modeling.\footnote{We recognize that such an approach does not account for potential behavioral dependencies between multiple interactions within the same day, which could provide additional insights into user preferences.}

%\hr{I don't this we should detail too much ...the next paragraph is way too apologetic...}An alternative approach is to order all interactions by day without considering their sequence within the same day. While this approach simplifies preprocessing, it introduces two major issues: it fails to maintain an accurate temporal order and neglects behavioral dependencies between items interacted with on the same day. In contrast, our method ensures at least a consistent temporal order, mitigating a critical inconsistency in the dataset and providing a more robust framework for sequence modeling.

%\hr{It doesn't sound really good to me...a set of adhoc settings...what did other works on sequential modeling did with Amazon data???...this can be major flaw for reviewers that are well familier with these datasets...} \gh{We talk about this point on the beginning of the project ... in the amazon dataset we don't have a resolution inside the day so you told me to sample item each day to create a real sequence. other works sort by the days and ingnore the problem if there is no a sequence at all or if the timeline is wrong}\bs{there are works that used various granularity levels like daily , weekly and monthly - some use the temporal information and some f=do not. you can see a survey on the properties of the studies of seq recommenders  - Nasir, Mahreen and Ezeife, C. I.. (2023). A Survey and Taxonomy of Sequential Recommender Systems for E-commerce Product Recommendation. SN Computer Science, 4 (6).
%https://scholar.uwindsor.ca/computersciencepub/59}

\begin{table}[bth]
\centering
\setlength{\tabcolsep}{1.0pt} % Adjust column spacing
\caption{Dataset statistics. For each domain, per number of interactions we also report the data density. Except for \texttt{Tools} dataset, all other domain dataset numbers are reported after user sampling is applied.} %\hr{Are those the actual numbers of our data after sampling??}\gh{yes, why?}}
\begin{tabular}{lcccr}
\hline
\textbf{Target Domain}       & \textbf{\# Users} & \textbf{\# Items} & \textbf{\# Interactions} & \textbf{Source Domains}\\
\hline
Electronics   & 26,319      & 12,031      & 165,541 \small{(5.2e-4)} &  Tools+Toys\\
Sports        & 19,244      & 10,036      & 120,122 \small{(6.2e-4)} & Toys+Electronics\\
Tools         & 10,962      & 5,820       & 66,164 \small{(1.0e-3)}  & Sports+Electronics\\
Toys          & 21,342      & 10,457      & 131,485 \small{(5.9e-4)} & Sports+Electronics\\
\hline
\end{tabular}
\label{tab:dataset_stats}
\end{table}

As shown in Table~\ref{tab:dataset_stats}, our datasets exhibit low density, which poses a significant challenge for traditional recommendation models. However, such data sparsity allows us to test our core hypothesis, that
language models that are trained in different domains can still be
used to provide recommendations in other, even different, domains. %\hr{I think this is a wrong statement for this work..our focus is on cross-domain and we only use a relatively "small" model. I would change the last sentence to: ``...such data sparsity allows to test our core hypothesis, that language models that were trained in different domains can be still used to provide recommendations in other, even different, domains''}\gh{Right, I changed this}. 
To balance computational efficiency with comprehensive evaluation, except for \texttt{Tools} domain, which is a relatively small dataset, we focus on a subset of users, selecting 40\% of users from the \texttt{Toys} and \texttt{Sports} domains and 30\% from the \texttt{Electronics} domain, ensuring  diverse and manageable datasets for analysis.

\subsubsection{Implementation}
%\hr{Guy add one or two sentence on the fact that you are using every time two source domains to train the third one...it is not clear what settings you used...e.g., for Electronics, which source domains you used, etc?} \hr{Also, what was the logic of selecting the source pair? e.g., if you used Tools+Sports to predict Toys, why not Electronics instead of Tools?..etc} \hr{All this info is currently missing} \gh{I added them in the end of this subsubsection, when we talk about X-Cross}
 To recall, we predict the next item using a \textbf{multiple-choice} task (see again Section~\ref{sec:multichoice task}). Accordingly, during both training and inference, for each next true item $i$, we sample $29$ negative items (resulting with 30 choices overall for the model to ``choose'' from). Following~\citep{sun2019bert4rec, lian2020personalized}, we obtain negative samples using a popularity-based sampling strategy. % to ensure robust evaluation. 
 During inference, we predict the next item (out of 30) as the one with the highest score.

As the backbone language model in our evaluation, we adopt the DeBERTa V3 base model~\citep{he2021deberta, he2021debertav3}. We make this language model choice  primarily to ensure a \textbf{fair comparison} with other cross-domain baseline models \citep{hou2022towards, hou2023learning, li2023text}, which have used similar model capacities (e.g., BERT) in their evaluation. %where we carefully selected this model based on its type and size, aligning with the constraints and capabilities of competing approaches. 
Additionally, we choose DeBERTa model as our base (language-model) encoder since it was also evaluated in the original LoRA paper~\citep{hu2022lora}. Therefore, such a choice provides a consistent and robust benchmark for assessing the effectiveness of X-Cross in cross-domain recommendation tasks.  To comply with DeBERTa's maximum sequence length (512 tokens), we truncate each item's title to a maximum of 8 words. % for efficient processing to the underlying language model (DeBERTa) maximum sequence length of 512 tokens.

We implement \textbf{X-Cross} using PyTorch and train the model for 40 epochs on two NVIDIA V100 GPUs, each with 32 GB of memory. The architecture utilizes \(n=2\) source domains (``experts''), where we use a holdout-set to select the top-two source domains with best zero-shot performance on the target domain. To enhance the model's efficiency and specialization, we only modify the top-9 layers of the source domain models. This decision is based on the understanding that the top layers of language models typically capture more abstract and high-level knowledge \citep{tenney2019bert}, making them better suited for cross-domain adaptation. Table~\ref{tab:dataset_stats} further details the source domains that we consider per each target domain. %\hr{Guy, we need some smart sentence to state why specific sources are selected??} \gh{What do you think about the assumption that performance on zero-shot settings is an indicator for cross-domain potential? } \hr{ah??..that's irrelevant here...why did you select specific two and not other combinations? per domain you had 6 possibilities!..a smart ass reviewer can say we cherry-picked source domains}

%Specifically, we pair the target domains with the following experts:
% \begin{itemize}
%     \item For \textbf{Tools}, we use experts trained on \textbf{Sports} and \textbf{Electronics}.
%     \item For \textbf{Sports}, we use experts trained on \textbf{Toys} and \textbf{Electronics}.
%     \item For \textbf{Toys}, we use experts trained on \textbf{Sports} and \textbf{Electronics}.
%     \item For \textbf{Electronics}, we use experts trained on \textbf{Tools} and \textbf{Toys}.
% \end{itemize}
% \hr{For reproducibility you must state which domains per target. By zero-shot you mean you evaluated them out-of-the-box? no fine-tuning?}\gh{yes, out-of-the-box e.g run tools on the sports test set and report the results}\gh{This better?}

We fine-tune the model using AdamW optimizer \citep{loshchilov2017decoupled}, configured with a learning rate of $5 \times 10^{-5}$, a weight decay of 0.01, and a batch size of 1 to address GPU memory constraints\footnote{We note that, each batch basically contains 30 prompts used to implement the model's multiple-choice task.}. We configure training hyperparameters using a holdout set, setting $\beta = 0.5$ and $\gamma = 0.4$ for optimal performance. Additionally, for LoRA fine-tuning, we employ a rank $r=16$ and a scaling factor $\alpha=32$, %\hr{Please cite papers} \gh{I cited LoRA paper, they get the lowest valid loss on with hyperparmeters. Do you think we need to tell that? I see that some papers didn't even mention the rank like tallrec and others only write the rank like llamarec} 
ensuring efficient parameter utilization while maintaining strong task performance~\citep{hu2022lora}. %Finally, similar to DeBERTa, we set 0.1 as the dropout parameter (Eq.~\ref{pooling}).

\subsubsection{Baselines} We compare X-Cross performance to several types of baselines.   %the performance of our model against a diverse set of baselines, encompassing domain-specific methods, merging strategies, cross-domain approaches, and traditional sequential recommendation models. 
We first evaluate several state-of-the-art ``traditional'' single-domain sequential recommendation models, namely: \textbf{SASRec}~\citep{kang2018self} -- a widely adopted self-attentive (causal) sequential recommendation model; \textbf{BERT4Rec}~\cite{sun2019bert4rec} -- a bidirectional transformer-based model which uses a masked learning approach; \textbf{FDSA} \citep{zhang2019feature}, which leverages self-attentive blocks to model item and feature transition patterns; \textbf{S$^3$-Rec} \citep{zhou2020s3}, which pre-trains the model to maximize mutual information for enhanced feature fusion; and lastly, \textbf{LLM-Rec} \citep{tang2023one}, a language model that represents both user history and the next item within a shared embedding space. While this model is generally designed for multi-domain settings, here we apply it to a single-domain setting due to the \textbf{lack of overlapping users across domains}. For this evaluation, we specifically use an encoder-only language model because it achieves the best results. Additionally, to ensure a fair comparison, we use the same encoder architecture (DeBERTa) as in our approach.

As a first-line of baselines for cross-domain recommendation, we fine-tune a domain-specific LoRA model~\citep{hu2022lora} for each \textbf{source} domain. Here we fine-tune each model independently on its respective (source) domain and evaluate it in a \textbf{zero-shot} setting on other domains. These set of baselines allows to assess the potential of transferability among different domains. %This differs from the other models, which we train and evaluated on each domain independently.

We next evaluate three state-of-the-art cross-domain sequential-recommendation models, namely: \textbf{UniSRec}, \textbf{VQ-Rec} and \textbf{RecFormer}. All these models require pretraining. For fair comparison, we pretrain these models on the exact source domains that we use for \textbf{X-Cross} and then fine-tune them on the target domain. \textbf{UniSRec}~\citep{hou2022towards} equips textual item representations with a mixture-of-experts (MoE)-enhanced adapter for domain fusion and adaptation; leveraging item-sequence and sequence-sequence contrastive learning tasks to pre-train transferable sequence representations. \textbf{VQ-Rec}~\citep{hou2023learning} learns vector-quantized item representations for transferable sequential recommendation~\citep{hou2023learning}, enabling effective representations across domains. %For this baseline, we use its pre-trained weights, and each time, fine-tune it on a given target domain\hr{This might change if you fine tune in advance our the same source domains}\gh{working now on the preptraining}\gh{I added this above}\gh{The parts I added ok? i did it for unisrec and reformer, only vqrec to go.}. 
% We note that, \textbf{VQ-Rec} cannot be evaluated on the \texttt{Tools} dataset due to insufficient data for the vector-quantization stage. 
Finally, \textbf{RecFormer}~\cite{li2023text} leverages language representations to model user preferences and item features, enabling effective next-item prediction, particularly in low-resource and cold-start scenarios.

As two competitive alternative baselines that also focus on integrating domain-specific knowledge, we consider \textbf{XLoRA}~\citep{buehler2024x} and \textbf{MeteoRA}~\citep{xu2024meteora}.  The \textbf{XLoRA} baseline incorporates core techniques from the XLoRA framework \citep{buehler2024x} into our setting. Specifically, we use embeddings from the last layer of the pre-trained model to scale all layer ``experts'', leveraging cross-layer attention to improve generalization across domains. Further adapting the \textbf{MeteoRA} framework \citep{xu2024meteora}, we implement a gating network for each layer. These networks dynamically integrate adapters by using embeddings from the pre-trained model as inputs, ensuring a more effective integration of domain-specific knowledge. Finally, we implement \textbf{Pooler + Scorer} baseline, which simplifies the architecture by fine-tuning only the pooler and scorer layers of the best-performing pre-trained model. We use this baseline to examine the necessity of integrating multiple domain-specific components. 

\subsubsection{Evaluation Metrics}
Following previous works~\citep{ren2024representation, yuan2020parameter, yuan2021one}, we divide each dataset into training, validation, and testing sets based on user splits, following a 3:1:1 ratio. We asses model performance using standard metrics for top-$k$ recommendation tasks. Specifically, we calculate Hit@$k$ for \( k = \{1, 3, 10\} \), and MRR for \( k = 10 \). Following \citep{kim2024large, zhou2020s3, kang2018self}, we conduct negative sampling during the evaluation phase in a manner consistent with the training process, ensuring reliable and comparable evaluation results. Finally, statistical significance of X-Cross performance is evaluated throughout using a two-tailed paired Student's t-test (\( p \leq 0.05 \)). %\hr{if you have some works to cite here, please do..} \gh{I add some works that use negative samples in evaluation}  %\bs{I don't understad this....} \gh{You want me to cite here paper that use negative sampling in the evaluation? }

\subsection{Overall Performance}
\label{sec:Results}
\begin{table*}[tbh]
\small
    \centering
    \renewcommand{\arraystretch}{0.8} % Adjust row height
    \setlength{\tabcolsep}{1.5pt} % Adjust column spacing
    \caption{Overall performance of X-Cross and baselines. Bold values denote the best performer. Superscripts $s$ and $i$ and subscript $c$ denote a (statistical) significant difference ($p\leq 0.05$) with the best single-domain, integrators and  the best cross-domain baselines, respectively.} %, evaluated using Hit@1, Hit@3, Hit@10, and MRR metrics.}
    \begin{tabular}{@{}lcccccccccccccccc@{}}
        \toprule
        \textbf{Model}                    & \multicolumn{4}{c}{\textbf{Tools}} & \multicolumn{4}{c}{\textbf{Sports}} & \multicolumn{4}{c}{\textbf{Toys}} & \multicolumn{4}{c}{\textbf{Electronics}} \\ 
        \cmidrule(lr){2-5} \cmidrule(lr){6-9} \cmidrule(lr){10-13} \cmidrule(lr){14-17}
                                  & Hit@1 & Hit@3 & Hit@10 & MRR & Hit@1 & Hit@3 & Hit@10 & MRR & Hit@1 & Hit@3 & Hit@10 & MRR & Hit@1 & Hit@3 & Hit@10 & MRR \\ \midrule
        \multicolumn{17}{c}{\textbf{Single Domain Recommendation}} \\ \midrule
        SASRec                   & 6.38  & 14.23 & 37.07 & 13.26 & 9.79 & 22.34 & 51.96 & 19.99 & 8.36 & 17.92 & 41.46 & 16.13 & 6.34 & 16.19 & 44.76 & 15.05 \\
        BERT4Rec  & 4.79  & 11.13 & 33.33 & 11.02 & 7.85 & 18.08 & 45.54 & 16.56 & 6.11 & 14.71 & 38.86 & 13.52 & 5.07 & 12.96 & 39.17 & 12.59 \\

        FDSA & 5.43  & 12.59 & 36.84 & 12.19 & 8.21 & 20.21 & 49.52 & 18.02 & 7.14 & 16.37 & 44.25 & 15.55 & 6.04 & 16.00 & 43.01 & 14.53 \\
        
        $S^3$-Rec & 5.15  & 11.67 & 35.02 & 11.56 & 8.68 & 20.21 & 50.58 & 18.46 & 6.75 & 16.44 & 41.51 & 14.72 & 5.62 & 14.67 & 43.69 & 13.98 \\

        LLM-Rec & 14.77  & 35.70 & 74.14 & 30.14 & 22.29 & 47.96 & 82.49 & 39.38 & 23.45 & 49.87 & 83.91 & 40.87 & 25.57 & \textbf{51.99} & \textbf{85.03} & 42.82 \\
        
        % SSE-PT                   & 9.17  & 20.94 & 46.39 & y & 12.49 & 25.79 & 53.41 & y & 10.97 & 23.11 & 49.8 & y & 9.49 & 20.85 & 47.64 & y \\ 
        \midrule
         \multicolumn{17}{c}{\textbf{LoRA Fine-Tuning on Source Domain / Zero-shot on Target Domain}} \\ \midrule
        Tools-LoRA              & 18.74 & 39.26 & 74.87 & 33.69 & 13.30 & 28.11 & 58.35 & 24.64 & 13.91 & 26.40 & 53.99 & 23.79 & 8.85 & 20.11 & 45.55 & 17.75 \\
        Sports-LoRA             & 12.40  & 25.58 & 55.95 & 22.74 & 27.98 & \textbf{51.49} & \textbf{82.62} & 43.49 & 14.10 & 25.49 & 49.50 & 22.92 & 8.49 & 17.63 & 44.03 & 16.48 \\
        Toys-LoRA               & 11.26 & 22.16 & 48.34 & 20.06 & 14.52 & 27.54 & 55.55 & 24.71 & 27.99 & 52.02 & 84.77 & 44.05 & 8.98 & 18.96 & 45.21 & 17.47 \\
        Electronics-LoRA        & 12.40  & 26.32 & 50.39 & 21.59 & 15.82 & 30.32 & 59.08 & 26.74 & 14.34 & 25.84 & 52.85 & 23.74 & 25.56 & 51.77 & 84.08 & 42.40 \\ 
        \midrule
        \multicolumn{17}{c}{\textbf{Cross-domain Recommendation}} \\ \midrule
        UniSRec    & 7.80  & 16.96 & 42.64 & 15.95 & 13.07 & 27.49 & 58.92 & 24.39 & 10.03 & 19.98 & 46.10 & 18.46 & 7.66 & 17.21 & 46.33 & 16.28 \\

        VQ-Rec & 6.84 & 15.55 & 41.04 & 14.42 & 10.81 & 24.06 & 55.18 & 21.58 & 8.03 & 17.36 & 43.15 & 16.08 & 7.09 & 16.36 & 43.84 & 15.34 \\

        RecFormer & 12.80 & 26.83 & 54.65 & 23.35 & 18.76 & 36.82 & 69.15 & 31.83 & 16.74 & 30.31 & 57.08 & 26.96 & 14.19 & 29.20 & 60.08 & 25.59 \\

        \midrule
        \multicolumn{17}{c}{\textbf{Domain Integration}} \\ \midrule
        XLoRA                  & 18.01 & 36.21 & 69.95 & 31.44 & 24.45 & 44.89 & 77.01 & 38.68 & 23.28 & 45.07 & 79.90 & 38.66 & 15.14 & 32.43 & 66.66 & 28.23 \\
        MeteoRA                  & 17.74 & 36.53 & 70.27 & 31.26 & 21.38 & 41.62 & 72.04 & 35.10 & 22.30 & 42.96 & 76.76 & 37.00 & 17.65 & 37.10 & 72.38 & 31.91 \\ 
        Pooler + Scorer      & 14.09 & 31.19 & 61.70  & 26.48 & 22.97 & 42.56 & 74.69 & 36.74 & 20.73 & 39.73 & 73.23 & 34.45 & 15.69 & 33.91 & 68.18 & 29.07 \\

        % \midrule
        % \multicolumn{17}{c}{Ours} \\ \midrule
        \midrule
        X-Cross (ours)                      &  \textbf{20.11}$_{\text{c}}^{\text{i,s}}$ & \textbf{40.36}$_{\text{c}}^{\text{i,s}}$ & \textbf{76.06}$_{\text{c}}^{\text{i,s}}$ & \textbf{34.87}$_{\text{c}}^{\text{i,s}}$ & \textbf{28.40} $_{\text{c}}^{\text{i,s}}$& 50.09$_{\text{c}}^{\text{i,s}}$ & 82.05$_{\text{c}}^{\text{i}}$& 43.36 $_{\text{c}}^{\text{i,s}}$ & \textbf{28.23}$_{\text{c}}^{\text{i,s}}$ & \textbf{52.66} $_{\text{c}}^{\text{i,s}}$& \textbf{85.66}$_{\text{c}}^{\text{i,s}}$ & \textbf{44.65}$_{\text{c}}^{\text{i,s}}$ & \textbf{26.25}$_{\text{c}}^{\text{i}}$ & 50.72$_{\text{c}}^{\text{i}}$ & 84.71$_{\text{c}}^{\text{i}}$ & \textbf{42.84}$_{\text{c}}^{\text{i}}$ \\ 
        
        \bottomrule
    \end{tabular}
    \label{tab:hit_mrr_comparison}
\end{table*}

We analyze the overall performance of \textbf{X-Cross} compared to the different types of baselines. First, examining the result of single-domain baselines, the \textbf{LLM-Rec} baseline demonstrates a significant performance improvement over the others. This showcase the benefit of leveraging embeddings derived from all item titles in the user's history. This holistic representation of user history enables \textbf{LLM-Rec} to outperform models that treat items' text independently.

Next, as we can observe, those baselines that we fine-tune with LoRA using a multiple-choice task on the target domain, perform even better  than \textbf{LLM-Rec}. Furthermore, the cross-domain potential becomes evident, as the baselines that we evaluate using a zero-shot setting also outperform most single-domain baselines. 

Next,  except for \textbf{LLM-Rec}, the cross-domain baselines achieve significantly better results than their single-domain counterparts (with \textbf{RecFormer} as the top performer). While these cross-domain baselines may not match LoRA fine-tuning's results for single-domain tasks, this does not necessarily reflect poorly on their cross-domain capabilities. Instead, it highlights the strength of holistic user history representation, as seen in both \textbf{LLM-Rec} and LoRA fine-tuning and also the potential of cross-domain recommendation among single-domain baselines.

We next examine the performance of the domain integration baselines, where we observe a mixed performance both among themselves and compared to other baselines. Specifically, in some cases, these baselines perform worse than the \textbf{Pooler + Scorer} baseline, which fine-tunes the best source domain's pooler and scorer. However, these integration approaches still provide a strong baseline, particularly when considering their efficiency. 

Finally, we first compare \textbf{X-Cross} side-by-side with those baselines that require a new LoRA adapter. Even though \textbf{X-Cross} requires significantly fewer parameters compared to training a new LoRA adapter, on \texttt{Sports} and \texttt{Toys} domains it almost reach the same accuracy of these baselines, while for \texttt{Tools} and \texttt{Electronics} domains it even outperforms these baselines. Furthermore, compared to alternative integrator-baselines, \textbf{X-Cross} exceed their performance by a large margin. All in all, these results serve as a strong evidence that \textbf{X-Cross} serves as both efficient and effective solution for cross-domain sequential recommendation tasks.

\begin{figure}[tbh]
    \centering
    \includegraphics[width=0.45\textwidth]{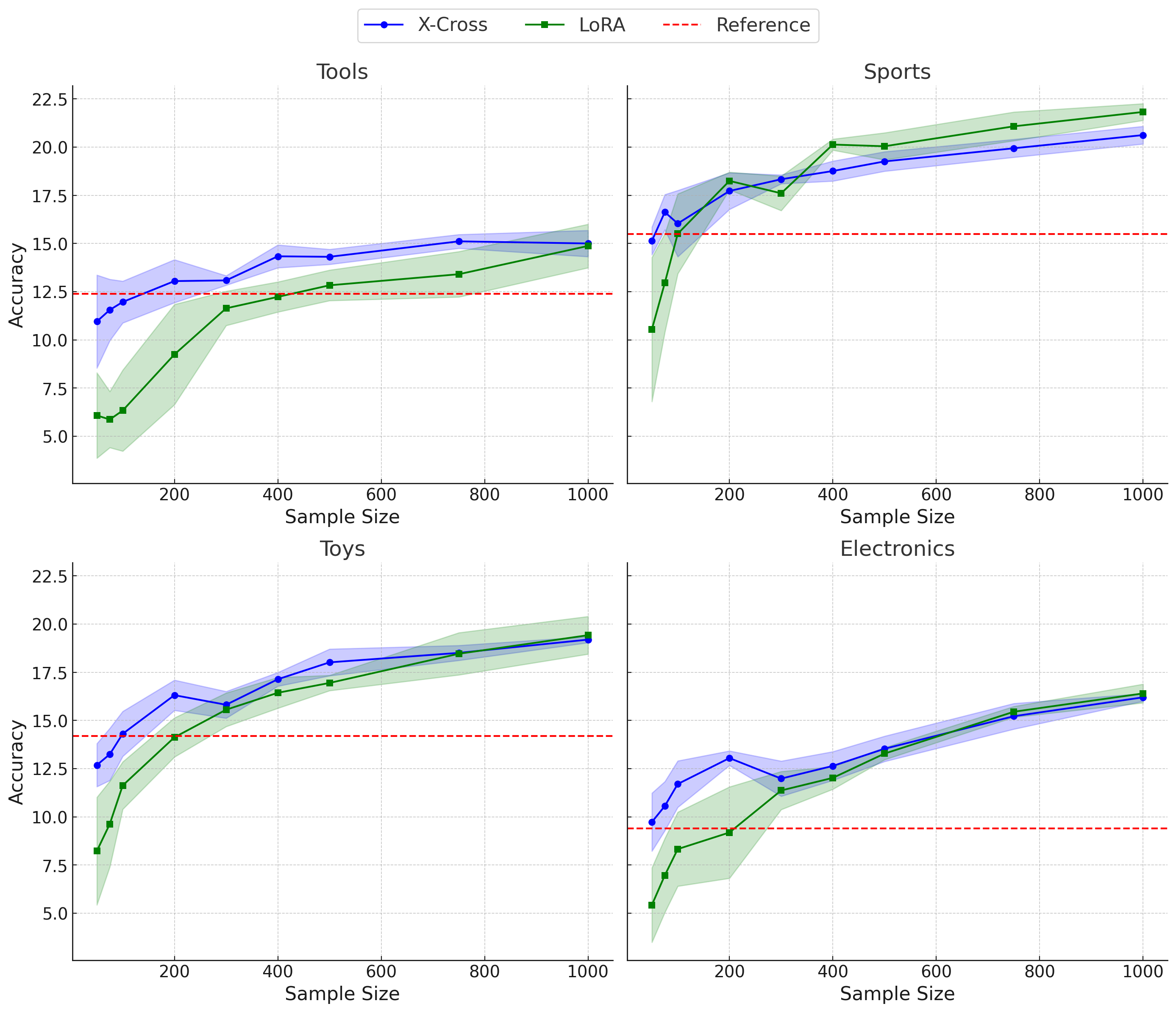}
    \caption{Accuracy (Hit@1) comparison across datasets for X-Cross and LoRA. The dashed red-line denotes the performance of the reference model.}
    \label{fig:accuracy_comparison}
\end{figure}

\begin{table}[tbh]
\centering
\scriptsize
\setlength{\tabcolsep}{1.5pt} % Adjust column spacing
\caption{Models performance under training data limitations.}
\label{tab:training_data_comparison}
\begin{tabular}{|l|c|c|c|c|}
\hline
\textbf{\begin{tabular}[c]{@{}l@{}}Target \\ Domain\end{tabular}} & \textbf{\begin{tabular}[c]{@{}l@{}}Significant for \\ $<$ M Samples\end{tabular}} & \textbf{\begin{tabular}[c]{@{}l@{}}X-Cross to Exceed\\ Baseline\end{tabular}} & \textbf{\begin{tabular}[c]{@{}l@{}}LoRA to Exceed\\ Baseline\end{tabular}} & \textbf{\begin{tabular}[c]{@{}l@{}}Gap (\%) \\ \end{tabular}} \\\hline
Tools          & 1000                                   & 200                              & 500                              & 60.0\%              \\ \hline
Sports         & 100                                    & 75                               & 200                              & 62.5\%            \\ \hline
Toys           & 300                                    & 100                              & 300                              & 66.7\%            \\ \hline
Electronics    & 300                                    & 50                               & 300                              & 83.3\%            \\ \hline
\end{tabular}
\end{table}

\begin{table*}[tbh]
    \centering
    \small
    \renewcommand{\arraystretch}{0.9} % Adjust row height
    \setlength{\tabcolsep}{1.5pt} % Adjust column spacing
    \caption{Ablation study results: Impact of removing key components on performance.} %Values represent Hit@1, Hit@3, Hit@10, and MRR@10 across the four domains.}
    \label{tab:ablation}
    \begin{tabular}{lcccccccccccccccc}
        \toprule
        \textbf{Variant} & \multicolumn{4}{c}{\textbf{Tools}} & \multicolumn{4}{c}{\textbf{Sports}} & \multicolumn{4}{c}{\textbf{Toys}} & \multicolumn{4}{c}{\textbf{Electronics}} \\
        \cmidrule(lr){2-5} \cmidrule(lr){6-9} \cmidrule(lr){10-13} \cmidrule(lr){14-17}
        & Hit@1 & Hit@3 & Hit@10 & MRR & Hit@1 & Hit@3 & Hit@10 & MRR & Hit@1 & Hit@3 & Hit@10 & MRR & Hit@1 & Hit@3 & Hit@10 & MRR \\
        \midrule
        X-Cross & 20.11 & 40.36 & 76.06 & 34.87 & 28.40 & 50.09 & 82.05 & 43.36 & 28.23 & 52.66 & 85.66 & 44.65 & 26.25 & 50.72 & 84.71 & 42.84 \\ 
        - Layers & 16.60 & 35.11 & 68.99 & 30.20 & 24.24 & 44.58 & 76.57 & 38.53 & 22.96 & 45.26 & 81.73 & 38.73 & 15.67 & 33.33 & 69.47 & 29.15 \\
        - Interactions & 17.46 & 38.49 & 75.01 & 32.62 & 26.73 & 49.55 & 81.84 & 42.10 & 26.10 & 50.97 & 83.35 & 42.51 & 23.02 & 46.12 & 80.72 & 38.95 \\
        - Experts & 19.33 & 39.67 & 73.78 & 33.68  & 27.57 & 49.75 & 81.55 & 42.62 & 25.81 & 49.03 & 83.98 & 41.80 & 26.06 & 50.06 & 82.88 & 42.13 \\
        \bottomrule
    \end{tabular}
\end{table*}
\subsection{Efficiency on Limited Training Data}
\label{sec: Efficiency}
We next wish to evaluate X-Cross performance under limited training data settings. To this end, for each target domain, we compare X-Cross side-by-side with a model that is fine-tuned with LoRA. Here, we remind that, using $n=2$ source domains, X-Cross requires only $25\%$ of the parameters of LoRA\footnote{Per each layer $l$, X-Cross learns the weights matrix $\mathbf{W}^{(l)}_{\text{concat}}$ which for $n=2$ translates to 4*2*768 total parameters; whereas LoRA learns two weight matrices $\mathbf{A}^{(l)}$ and $\mathbf{B}^{(l)}$, together having  2*16*768 total parameters.}. Yet, as we shall shortly demonstrate, despite this modest decrease in parameter size, X-Cross achieves competitive results against LoRA in terms of performance, underscoring its effectiveness.

We evaluate both models (X-Cross and LoRA) using training datasets of varying sizes: \{50, 75, 100, 200, 300, 400, 500, 750, 1000\}. For each training dataset size, we randomly sample five distinct subsets of the specified size from the full dataset. We train both models separately on each of these subsets.
As a reference model for minimum performance requirement, for each target domain, we consider the source model with best zero-shot performance which is trained on one of the other (source) domains (e.g., for \texttt{Tools}, the reference model is LoRA-fine-tuned either on the \texttt{Sports}, \texttt{Toys} or \texttt{Electronics} domain). 

We describe our results in Figure~\ref{fig:accuracy_comparison}, where we evaluate all three models on the full test-set to identify the minimum amount of training data required for either LoRA or X-Cross to surpass the performance of the reference model (further represented by the red dashed line). %Overall, the goal of this evaluation is to highlight the robustness of X-Cross under limited training data settings. 
% Moreover, statistical significance testing using a two-tailed paired Student's t-test (\( p \leq 0.05 \)) demonstrated that our model consistently outperformed the standard LoRA model across the evaluated training set sizes, confirming the reliability and generalizability of its performance.
We further summarize the training data requirements across domains of both models in Table~\ref{tab:training_data_comparison}. The first column indicates the maximum number of samples (\( M \)) in the training set for which X-Cross performance remains statistically significantly better than LoRA. The second and third columns represent the minimum number of training samples required for X-Cross and the standard LoRA adapter to exceed the reference model, respectively. The final column, ``Gap (\%)'', shows the percentage reduction in the amount of training data required by X-Cross compared to LoRA. %, calculated as \( \frac{\text{LoRA Samples} - \text{Our Model Samples}}{\text{LoRA Samples}} \times 100 \).
Further note that, the values in Table~\ref{tab:training_data_comparison} represent the \textbf{smallest} sampled training dataset sizes at which the performance threshold is exceeded. If the threshold is not surpassed at a given size (e.g., 100) but is at the next increment (e.g., 200), the latter value (200) is recorded in the table to reflect this milestone in the sampling process.
Overall, these empirical results indicate that X-Cross attains a steeper initial learning curve, reaching to above-reference performance with substantially fewer samples.

\subsection{Impact of Number of Model Layers}
\label{sec: Layers}
\begin{figure}[th]
    \centering
    \includegraphics[width=0.3\textwidth]{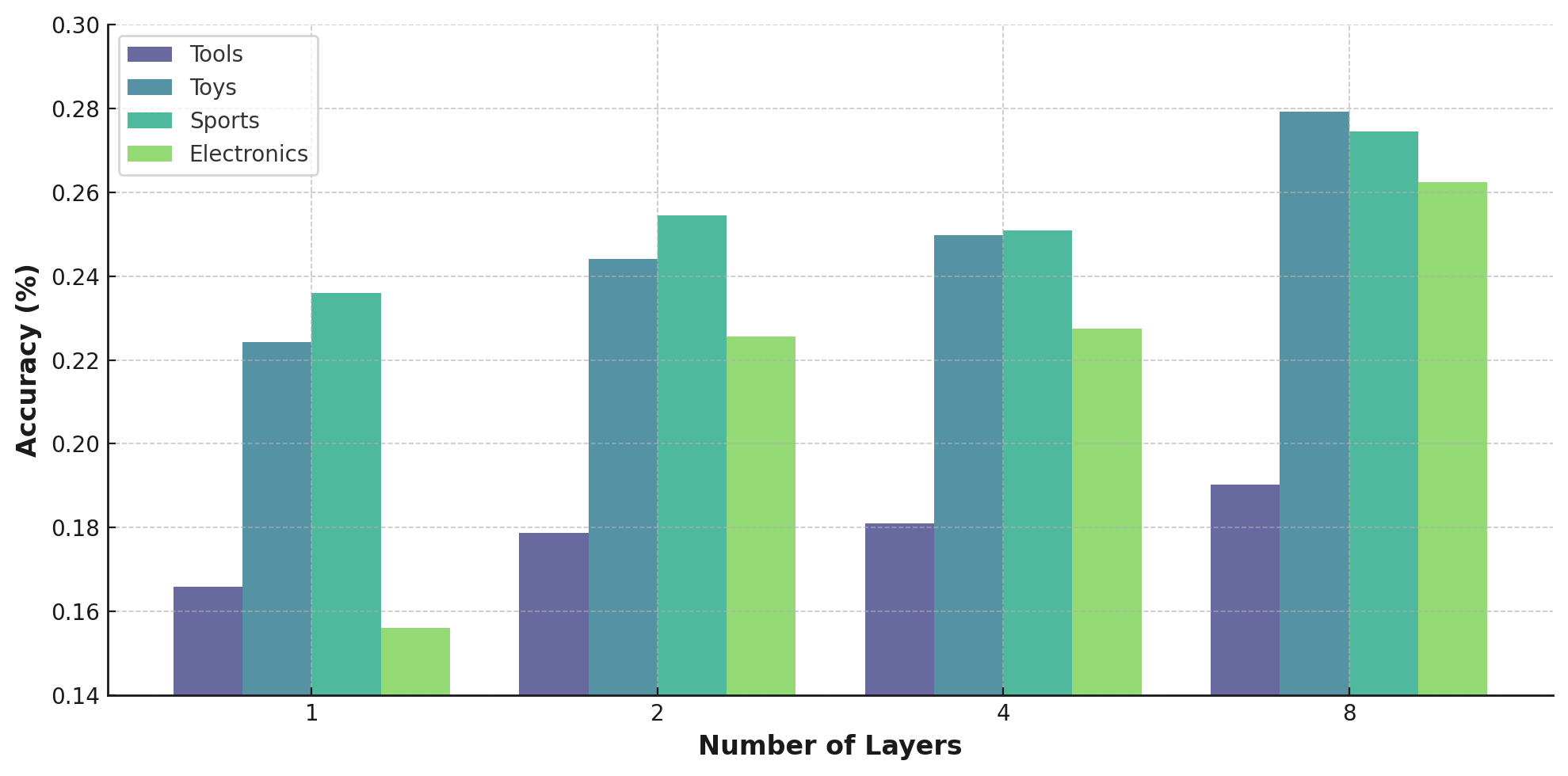}
    \caption{Accuracy (Hit@1) vs number of layers.}
    \label{fig:layers}
\end{figure}

We further investigate the contribution of individual layers within X-Cross. To this end, we vary the number of layers integrated into the model. Specifically, we evaluate its performance when scaling and integrating either 1, 2, 4, or 8 layers during the first three stages, starting from the \textbf{topmost} layer and down. The results in Figure~\ref{fig:layers} show that, while adding layers generally improves recommendation accuracy, the performance gains are not uniform. Some layers capture more domain-relevant knowledge and significantly impact performance, while others contribute less.

\subsection{Ablation Study}
\label{sec:ablation}
We next perform an ablation study to evaluate the  relative contributions of the key components of X-Cross. %, we performed an ablation study on the four domains: Tools, Sports, Toys, and Electronics. 
For that, each time, we systematically alter the full implementation of X-Cross by removing a specific component and measuring its impact on performance. We summarize the results of the ablation study in Table~\ref{tab:ablation}.
First, by setting both $\beta=0$ and $\gamma=0$ (in Eq.~\ref{domain intergation}) we remove the dynamic integration mechanism between source domains at each layer, allowing each one to operate independently and combining their outputs only at the \textbf{final stage} (Eq.~\ref{final representation}). As we can observe, this simplification (denoted ``\textbf{-Layers}'' in Table~\ref{tab:ablation}) which eliminates layer-wise collaboration, leads to a significant (and largest) performance drop, emphasizing the critical role of dynamic inter-layer integration. 

Next, we exclude the interactions component during the integration process by setting  \(\gamma = 0\) while still keeping $\beta=0.5$ (denoted ``\textbf{-Interactions}'' in Table~\ref{tab:ablation}). This modification, which disables the interaction terms \((\mathbf{h}_{m}^{(l)} - \mathbf{h}_{m'}^{(l)})\), reduces the model's ability to capture nuanced inter-domain relationships, resulting in a notable performance decline. 

Lastly, we eliminate the contributions of other source domains entirely by setting \(\beta = 0\) while still keeping $\gamma=0.4$ (denoted ``\textbf{-Experts}'' in Table~\ref{tab:ablation}). This modification leaves the domain-specific model and the interactions of the models active for each input. As we can observe, removing this functionality from X-Cross results in significant performance drop as well, which attests again to the importance of considering the contributions of other domains.

\subsection{Domain Convergence Analysis}
\label{sec: Convergence}
We conclude this section by investigating why some domains converge faster than others during training. Referring back to Figure \ref{fig:accuracy_comparison}, using five random samples of 1000 training examples from each domain's dataset, we observe significant differences in convergence performance, with an average accuracy of $21.82$, $19.42$, $16.40$ and $14.87$ for \texttt{Sports}, \texttt{Toys}, \texttt{Electronics} and \texttt{Tools} domains, respectively. Despite all domains being trained on identical sample sizes, the disparities in performance are pronounced. Interestingly, the bottom two domains, \texttt{Tools} and \texttt{Electronics}, represent extremes in dataset size -- with \texttt{Tools} being the smallest and \texttt{Electronics} the largest. This contradiction prompted us to further investigate which possible factors are influencing domain convergence, such as inherent dataset characteristics or domain-specific complexities.

Our initial attempts to explain these differences using classic features of recommendation datasets, such as the number of unique users and items, dataset density, average interactions per user, or per-item distributions, appear to be inconclusive. Similarly, analyzing the characteristics of pre-trained embeddings of individual items or of the entire history of the user offer no clear explanation. However, we do observe that the embedding cosine similarities within domains are exceptionally high, averaging \textbf{0.95}, suggesting significant homogeneity in these domains’ representations. This drives us even further to investigate prompt-specific features. More specifically, we focus on two main prompt properties: 1. \textit{prompt length} (measured by mean length) and 2. \textit{prompt diversity} (measured as standard deviation of the length). To this end, we train a simple linear regression model with the dependent parameter as the model's accuracy and the regressors are both \textit{prompt length} and \textit{prompt diversity}. Overall, the two prompt properties explain $86\%$ of the variance of the model's accuracy, demonstrating strong relationship between prompt quality and performance. Moreover, based on the regression coefficients, \textit{prompt length} exhibit negative relationship (-0.65) with model accuracy, whereas \textit{prompt diversity} has a positive one (1.60). This suggests that fine-tuning the model with shorter and more diverse prompts results in better accuracy.

\section{Conclusion}
We have introduced X-Cross, a novel cross-domain sequential recommendation model that dynamically integrates multiple language models at both the layer and input (sample) levels. %We have demonstrated that 
X-Cross achieves superior performance compared to state-of-the-art single-domain and cross-domain baselines while remaining highly efficient, outperforming parameter-efficient fine-tuning (PEFT) methods like LoRA, particularly in limited training scenarios. Moreover, %Additionally, 
X-Cross introduces a new approach to integrating LoRA adapters across domains, yielding improvements in cross-domain sequential recommendation tasks that highlight its adaptability and robustness.

As future work, we wish to explore the relationship between source and target domains to better understand how domain selection impacts performance, paving the way for further improvements in cross-domain recommendation systems.

\balance
\bibliographystyle{ACM-Reference-Format}
\bibliography{references}
\end{document}